\documentclass[10pt, conference]{IEEEtran}

\usepackage[noadjust]{cite}

\usepackage{booktabs}
\usepackage{graphicx}
\usepackage{amsmath}
\usepackage{longtable}
\usepackage{latexsym}
\usepackage{times}
\usepackage[english]{babel}
\usepackage[utf8]{inputenc}
\usepackage[T1]{fontenc}
\usepackage{float}
\usepackage{enumerate}
\usepackage{setspace,ifthen}
\usepackage{amsthm} 
\usepackage{amsmath}
\usepackage{amssymb}
\usepackage[usenames,dvipsnames]{xcolor}
\usepackage{mathrsfs}
\usepackage{units}
\usepackage{braket}
\usepackage{upgreek}
\usepackage{hyperref}

\usepackage{listings}

\definecolor{codegreen}{rgb}{0,0.6,0}
\definecolor{codegray}{rgb}{0.5,0.5,0.5}
\definecolor{codered}{rgb}{0.8,0,0}
\definecolor{backcolour}{rgb}{0.95,0.95,0.92}
\definecolor{commentcolor}{rgb}{0.2, 0.5, 0.6}

\lstdefinestyle{mystyle}{
    backgroundcolor=\color{lightgrey},   
    commentstyle=\color{commentcolor}\itshape,
    keywordstyle=\color{codegreen}\bfseries,
    numberstyle=\tiny\color{codegray},
    stringstyle=\color{codered},
    basicstyle=\ttfamily\scriptsize,
    breakatwhitespace=true,         
    breaklines=true,                 
    captionpos=b,                    
    keepspaces=true,                 
    numbers=none,                    
    numbersep=5pt,                  
    showspaces=false,                
    showstringspaces=false,
    showtabs=false,                  
    tabsize=2,
    postbreak=\mbox{\textcolor{gray}{$\hookrightarrow$}\space},
    framexleftmargin=5pt,
    framextopmargin=0pt,
    framexbottommargin=0pt,
    frame=tb, framerule=0pt
}

\lstdefinelanguage{mypython}[]{Python}{
    morekeywords={as}
}

\lstdefinelanguage{XML}
{
  morestring=[b]",
  morecomment=[s]{<?}{?>},
  stringstyle=\color{codered},
  identifierstyle=\color{codegreen}\bfseries,
  keywordstyle=\color{blue},
  morekeywords={xmlns,version,type,name}
}

\usepackage{caption}
\usepackage{tipa}
\definecolor{hyperlink_blue}{RGB}{0, 105, 171}
\definecolor{citation_blue}{RGB}{120, 2, 199}
\definecolor{url_blue}{RGB}{2, 28, 196}
\definecolor{lightgrey}{RGB}{240, 240, 240}

\newcommand{\code}[1]{\texttt{#1}}

\hypersetup{
final=true,
plainpages=false,
pdfstartview=FitV,
pdftoolbar=true,
pdfmenubar=true,
bookmarksopen=true,
bookmarksnumbered=true,
breaklinks=true,
linktocpage,
colorlinks=true,
linkcolor=hyperlink_blue,
urlcolor=url_blue,
citecolor=citation_blue,
anchorcolor=black
}

\addto\extrasenglish{%
}
\addto\extrasenglish{%
}

\makeatletter
\renewcommand{\p@subsection}{}
\renewcommand{\p@subsubsection}{}
\makeatother

\title{Programming the full stack of an open-access quantum computer}

\date{\today}

\author{\IEEEauthorblockN{Virginia Frey\IEEEauthorrefmark{2}$^{1,2}$, Richard Rademacher\IEEEauthorrefmark{2}$^{1,2}$, Elijah Durso-Sabina$^{1,2}$, Noah Greenberg$^{1,2}$,\\ Nikolay Videnov$^{1,2}$, Matthew L. Day$^{1,2}$, Rajibul Islam$^{1,2}$ and Crystal Senko$^{1,2}$}

\IEEEauthorblockA{$^{1}$Department of Physics and Astronomy, University of Waterloo, Waterloo, N2L 3R1, Canada\\
$^{2}$Institute for Quantum Computing,
University of Waterloo, Waterloo N2L 3R1, Canada}}

\begin{document}

\maketitle
\thispagestyle{plain}
\pagestyle{plain}

\begingroup\renewcommand\thefootnote{\IEEEauthorrefmark{2}}
\footnotetext{These two authors contributed equally to this work.}
\endgroup

\begingroup\renewcommand\thefootnote{\hspace{2mm}}
\footnotetext{Contact \href{mailto:virginia.frey@uwaterloo.ca}{virginia.frey@uwaterloo.ca}}
\endgroup

\begin{abstract}

We present a new quantum programming language called `Quala' that enables true full-stack programming of quantum hardware. Quala allows seamless integration of abstraction layers such as the digital circuit layer and the analog control pulse waveform layer. Additionally, the language supports user-issued low-level hardware instructions like FPGA actions. Mid-circuit measurements and branching decision logic support real-time, adaptive programs. This flexibility allows users to write code for everything from quantum error correction to analog quantum simulation. The combination of a user-facing calibration database and a powerful symbolic algebra framework provides users with an unprecedented level of expressiveness and transparency. We display the salient characteristics of the language structure and describe how the accompanying compiler can translate programs written in any abstraction layer into precisely timed hardware commands. We intend for this language to bridge the gap between circuit-level programming and physical operations on real hardware while maintaining full transparency in each level of the stack. This eliminates the need for ``behind-the-scenes'' compilation and provides users with insights into the day-to-day calibration routines.

\end{abstract}



\section{Introduction}

Over the past decade we have seen an explosion of quantum programming languages (QPLs) and quantum programming paradigms \cite{quantum-languages-overview, heim2020quantum} that enable users to interact with quantum hardware and implement quantum algorithms. To date, most QPL development effort has been focused on circuit-level description languages \cite{qasm, cross2021openqasm, cirq, jaqal, pyquil}, which allow users to express quantum algorithms in the standard quantum circuit picture consisting of a series of time-independent single- and multi-qubit gates \cite{nielsen2002quantum}. 

However, the level of abstraction that this circuit layer offers is limited in its usefulness as long as digital quantum computing is dominated by errors. Gate fidelities must be improved by orders of magnitude to reach the thresholds for error correcting codes small enough to be run on current quantum computers \cite{knill1998resilient}. To get us closer to that goal, using quantum computers as the analog machines that they really are may help realize the necessary operational fidelity for the highly anticipated breakthrough in quantum computing \cite{shi2020resource, jordan2008quantum}. Analog control pulse engineering for example has seen broad adoption across various hardware platforms with encouraging results \cite{roos2008ion, milne2020phase, khaneja2005optimal, goswami2003optical, shapira2018robust}, suggesting that control pulse design will remain a productive area of research for some time. Apart from general-purpose quantum computing, there are several promising near-term applications (\emph{e.g.} analog quantum simulation \cite{monroe2021programmable}, digital-analog quantum computation \cite{parra2020digital}, simulation \cite{rajabi2019dynamical}) that also require a level of control over the quantum hardware that goes far beyond the digital circuit layer.


A ``full-stack'' quantum programming language, that integrates all the required low-level analog hardware controls seamlessly with digital circuit abstraction layers would be ideal for these applications. Though promising candidates for ``pulse-level'' control have been proposed \cite{qiskit-openpulse, cross2021openqasm}, they do not give users control over hardware controls beyond the level of control and readout waveforms. However, the ability to access hardware controls beyond simple waveform generation opens up the possibility for much broader fields of research,  as transport-based quantum control in the trapped ion QCCD architecture \cite{pino2021demonstration}. Additionally, in the case of remote-access quantum hardware, it reduces the need for ``behind-the-scenes'' compilation and allows for full transparency in not only how abstract gates are implemented on the hardware, but also in the calibration processes that are required to tune up the whole system.

These calibration processes pose an essential challenge in quantum hardware platforms. The tune-up routines to achieve peak device performance are generally not modifiable or even exposed to the user since they require full access to the experimental hardware and a language that is expressive enough to describe this level of access. 

To address these challenges, we have developed a ``true'' full-stack programming language that provides the complete connection between the high-level circuit layer and the low-level timing commands for FPGA hardware. Quala also addresses the calibration challenge in a transparent way. A powerful run-time decision logic allows conditional execution directly in the user's code. A reusable library of standard gates allows algorithm designers to focus on high-level circuits.  We designed this language to run the the QuantumION platform, which is a remote-access trapped ion quantum processor built at the Institute for Quantum Computing in Waterloo, Canada\footnote{A separate publication about the hardware platform is forthcoming, and we will thus keep its description here brief.}. Our programming language fully describes \emph{all} of the operations required to realize a trapped-ion quantum computer. While the accompanying hardware setup for our experiment is tailored to the trapped-ion architecture, the concepts and ideas that went into our language design are generic enough to allow portability to other experimental setups. The language even supports other implementations of quantum computing so long as they are controlled with FPGA hardware.

Many academic as well as industrial labs are facing problems such as the transition between the circuit and the timing layer, real-time decision logic, and the incorporation of calibration data with symbolic algebra. In this manuscript we present an overview of this generic quantum computing programming language and discuss our proposed solutions to these common experimental problems. This manuscript is structured as follows: We begin with an outline of the design philosophy and principles that we set out to achieve with this language. Next, we articulate the core requirements that are necessary to unambiguously specify quantum programs more broadly. Following this, we describe the three elements that we have identified as key abstractions to organize a fully-specified quantum program. 
After this, we continue with an in-depth discussion of the formal structure of the language and the individual language elements that make up a full program. We discuss the compilation process of programs written at any layer of the stack and provide references for integration of the language with FPGA hardware. We conclude our work with a brief summary of the highlights of the language and make recommendations for user adoptions and future development work.



\section{The Quala Programming Language}

Quala (pronounced \textipa{/'kw{\"a}l@/}) is a true full-stack programming language designed originally in the context of the QuantumION platform. While the name is inspired by the project itself (\emph{QUA}ntumion \emph{LA}nguage), the design principles and suggestions we put forth here range far beyond this specific platform. It is a meta-language that we have defined in XML, and users can interact with it through various language bindings in languages such as Python, Matlab and Julia. Before we go into the detailed implementation of the language, we begin with a discussion of our design philosophy.

\subsection{Background and design philosophy}

Our design philosophy behind Quala is to create a programming interface that is fully transparent to its users, \emph{i.e.} with no ``behind-the-scenes'' compilation of input circuits or hidden parameters and calibration routines that may yield unexpected results. To achieve this, we have designed our language according to the following principles:

\begin{itemize}
    \item \textbf{Full-stack control.} Users can program at all layers of the programming stack from hardware-agnostic, high-level quantum circuits all the way to precisely timed hardware commands. The transition between our \emph{Gate Layer} and \emph{Timing Layer} is seamless; it is easy to switch between these two complementary views of a quantum program (even within the same program).
    \item \textbf{Transparent Calibration} All calibration data are global to the machine and stored in a historical database. Calibration programs are written in the same language and users can inspect every routine.
    \item \textbf{Controllability over simplicity.} We sacrifice the simple circuit-level programming picture and replace it with an interface that is more complex to enable more expressiveness.
    \item \textbf{Open source design.} We go beyond simply publishing the source code\footnote{With the exception of proprietary information about selected hardware drivers.}, and embrace the spirit of open-source design by exposing the design of gates, timing functions, and the calibration operations within the language itself.  
\end{itemize}

The most striking consequence of these design principles is that our language not only supports programming of hardware-agnostic quantum circuits but enables a framework for much broader areas of research. Quala  can be used for expressing control pulse design problems, analog quantum simulation, and, in the context of our platform, even fundamental ion trap research in a QCCD architecture. In the next section, we describe the features that support all these experimental problems.

\subsection{Unique requirements of quantum programs}
The following features are required to span the space of possible quantum computing experiments:

\begin{itemize}
    \item \textbf{Support for high-level quantum circuits.} In the quantum circuit picture, quantum information is manipulated by applying temporal sequences of quantum gates to a register of qubits. 
    This abstraction assumes that the key information is the \textit{order} in which quantum gates are applied, not the time at which they occur. Thus, a QPL must support specifications that look like a sequence of quantum gates (perhaps applied in parallel) with no explicit timing information beyond "which gate comes after which". We refer to this programming paradigm as our Gate Layer.
    \item \textbf{Support for precision-timed events.} As previously discussed, quantum hardware is highly dependent on the precise details of which operations are applied at which times. For example, a quantum gate composed of a laser pulse will be highly error prone if the laser is turned off 100 nanoseconds too early. Thus, a QPL capable of specifying all the details of what occurs "under the hood" must have a consistent mechanism to specify the time at which different experimental parameters (such as laser amplitudes, trap voltages, etc) are changed. Moreover, these specification mechanisms must integrate seamlessly with Gate Layer programming if both are to be realized within the same language. We refer to this programming paradigm as our Timing Layer.
    \item \textbf{Custom waveform control.} Beyond simply specifying the time at which a quantum gate or other operation occurs, it is necessary to specify full details of the control pulses to maximize system performance. For example, a square laser pulse may have much worse performance as a quantum gate than a laser pulse with custom amplitude shapes \cite{choi2014optimal}. Our language can support custom waveforms in multiple ways, from the direct application of Timing Layer controls to build up a detailed description of a waveform (e.g. by specifying a different voltage change at each time step), to allowing users to provide and play back their own waveforms, to allowing control of custom digital hardware (such as digital oscillators and interpolators) that simplify the parametrization of the waveforms.
    \item \textbf{Real-time decision logic.} Many quantum programs benefit from the ability to measure a subset of qubits and subsequently perform \textit{different} operations based on the outcome of the qubit measurements. Our language provides a generic mechanism to branch between different, reusable segments of code.
    
\end{itemize}
Having listed the core requirements the language must satisfy to support all of our envisaged applications, we now turn our attention to key enabling features we have designed to aid in the usability and modularity of specifying time sequences for an arbitrarily large number of physical voltage channels that are used to orchestrate a quantum program.


\subsection{Key technical features}

\noindent
Our language must implement transparent, full-stack control without incorporating hardware specific commands into the integral structure of the language. Users must be able to access all valid control parameters and calibration values necessary to run experiments. To this end, our language framework incorporates the following key technical features:

\begin{enumerate}
    \item A \emph{Calibration Database}, to store and manage all experimental parameters relevant to user programs
    \item A \emph{Symbolic Algebra} framework that allows for flexible integration of calibration parameter into user programs
    \item A \emph{Standard Library} that contains all relevant experimental subroutines to facilitate the programming process. Standard gates are part of this library, and are built up of fundamental, Timing Layer functions utilizing parameters from the Calibration Database.
\end{enumerate}

\noindent
While the information in the Calibration Database and the Standard Library is hardware specific, the concepts are hardware-agnostic. In the following sections, we use example parameters applicable to our trapped-ion system to showcase how these features work together. Similar examples could be conceived with parameters and routines for different types of quantum hardware.

\subsubsection{Calibration Database}




The Calibration Database contains a large collection of machine parameters that users may require to run their experiments. These include both parameters which are actively calibrated and parameters which remain constant. Parameters in the first category could include laser beam intensities and alignments, pulse durations and amplitudes, as well as higher-level parameters such as SPAM (state preparation and measurement) errors and gate fidelities. Parameters in the second category are not actively calibrated but may be relevant to different calculations that include calibration parameters. 
These parameters include physical constants (\emph{e.g.} ion energy level splittings) and experimental constants (\emph{e.g.} the direct digital synthesis (DDS) sample clock frequency).

Parameters that are calibrated regularly are stored with associated calibration dates and times, and users have access to the entire calibration history of those values. To use a calibration parameter within a program, we provide language constructs like the \code{NamedConstant} directive. In the Python binding to our language, users can access parameters as follows:


    
\begin{lstlisting}[language=mypython]
import quala as ql

ql.NamedConstant("RamanRedSidebandFrequency", date="most-recent")
\end{lstlisting}

When used within a program, the language compiler (see \autoref{sec::compiler}) will insert the corresponding value at the time of compilation. Until then, the expression remains symbolic. This is particularly useful in the context of shared-access quantum computers, where several users might submit their programs to a queue:  when calibration processes in the background update certain parameters while the user's program is still in the queue, the compiler will automatically insert the most recent calibration value that is available at the time the program is actually run.

At any point, the user can also inspect the values before compiling their program through a dedicated call to the database:


\begin{lstlisting}[language=mypython]
>>> ql.query_database("DefaultMicrowaveRabiRate", date="most-recent")
.. <DatabaseEntry name="DefaultMicrowaveRabiRate" value="1" units="MHz" date="2021-05-31-08-55">
\end{lstlisting}

\subsubsection{Symbolic Algebra}

The Calibration Database alone is not sufficient to allow quantum programs that behave predictably over variations in machine parameters. Calculating every permutation would be inefficient and calibrations would likely conflict with each other. Instead, the Calibration Database collects only critical parameters that can be combined mathematically. We developed a Symbolic Algebra as an integral part of our language to allow for this.

In the binding languages, the Symbolic Algebra is enabled through overloading the standard algebraic operators. For example, given the example parameter for a microwave Rabi rate from above, we can calculate the corresponding time it takes to perform a $\pi$-pulse via:

\begin{lstlisting}[language=mypython]
pi_time = 3.14159 / ql.NamedConstant("DefaultMicrowaveRabiRate")
\end{lstlisting}

\noindent
In the actual XML language, this expression would be written as:



\begin{lstlisting}[language=XML]
<qi:DivisionOperator>
  <qi:NumericLiteral>3.14159</qi:NumericLiteral>
  <qi:NamedConstant name="DefaultMicrowaveRabiRate">
</qi:DivisionOperator>
\end{lstlisting}

Our language supports all standard algebraic operations as well as basic boolean logic.

\subsubsection{Standard Library}

We combine the Calibration Database and the Symbolic Algebra in what we call the Standard Library: a set of pre-defined \emph{calculations}, time-dependent \emph{functions} and time-independent \emph{gates}. For example, the $\pi$-time from above is available through the \code{NamedCalculation} directive:

\begin{lstlisting}[language=mypython]
ql.NamedCalculation("DefaultMicrowavePiTime")
\end{lstlisting}

 Standard experimental routines that are naturally described in the Timing Layer, such as the laser cooling of ions and shuttling routines, are available in the Standard Library through the \code{FunctionCall} directive:

\begin{lstlisting}[language=mypython]
ql.FunctionCall("DopplerCooling", duration=ql.NumericLiteral(3, "ms"))
\end{lstlisting}

\noindent
Quantum Gates, unlike timing functions, focus on the structural connection of information qubits rather than precisely timed commands. For example, the two-qubit controlled-not (CNOT) gate can be called using the \code{GateCall} directive:


\begin{lstlisting}[language=mypython]
ql.GateCall("CNOT", qubits=[0, 1])
\end{lstlisting}

\noindent
Named calculations, functions and gates form the whole of the Standard Library. All definitions within it exist on the user's computer and can be copied, used as-is, or used as a template for custom user functions and gates.


Standard Library procedures used within a program are automatically included in the program Header (see \autoref{sec::program_components}). There is no further modification or addition to these routines on the compiler side. The user program thus contains the \emph{entire} code required to carry out the experiment.

\subsection{Demonstration of full-stack control}

The Calibration Database, the Symbolic Algebra and the Standard Library are the core features of our language. Taken together, they enable users to program our machine at an abstraction level of their choosing and fully customize any of the intermediate operations.

\begin{figure}[t!]
\includegraphics[scale=1.0]{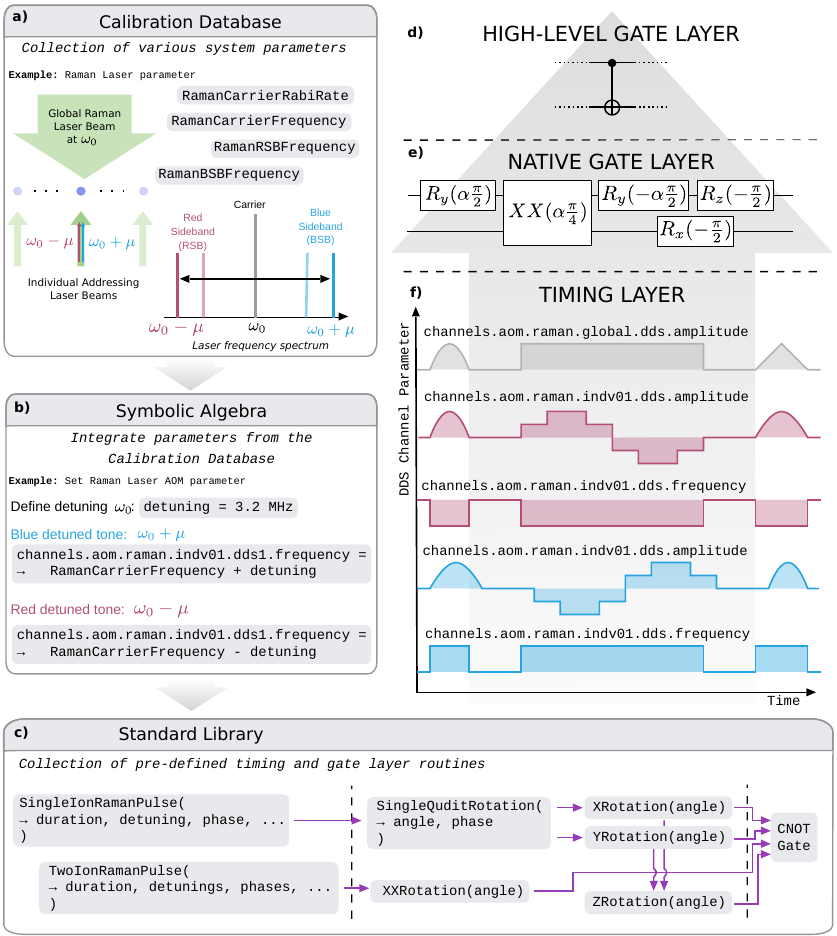}
\caption{Features of the Quala language illustrated through the ``top-down'' reconstruction of a CNOT gate. \textbf{a} The Calibration Database stores all parameters relevant for the experiment. Insets show schematics of the Raman laser beam configuration with laser frequency configurations required for single- and two-qubit interactions.
\textbf{b} The Symbolic Algebra provides the framework through which calibration parameters can be incorporated into programs. Parameters are referred to by name and can be used in standard algebraic expressions. \textbf{c} The Standard Library uses the Symbolic Algebra framework to define a set of re-usable experimental routines such as Timing Layer functions and high-level gates. The frequency configurations depicted in \textbf{a} and \textbf{b} are absorbed into functions for single- and two-ion laser pulses which form the basis for single- and two-qubit gates in the native Gate Layer. The flowchart depicts all functions and native gates required to realize a CNOT gate between two qubits. \textbf{d-f} show a schematic breakdown of the CNOT gate into native gates and individual Timing Layer instructions- illustrated through amplitude and frequency profile of selected DDS channels in the setup. 
}
\label{fig::cnot}
\end{figure}

This interplay is illustrated in \autoref{fig::cnot} using the ``bottom-up'' construction of a CNOT gate \cite{nielsen2002quantum} as an example. Fundamentally, a CNOT gate (or any ``gate'', for that matter) is an abstract operation that does not correspond directly to a single experimental action, but instead to a series of precisely timed operations. For laser-based gates in our ion trap qubits, these operations are implemented through laser-atom interactions \cite{molmer1999multiparticle, sackett2000experimental}. Our language allows users to specify all relevant parameters for this laser-ion interaction, from pulse amplitude and frequency profiles to phases and even beam pointing and polarization control. This control allows users to generate pulses with  custom amplitude, phase, or frequency modulation. These techniques have all been shown to greatly increase gate fidelitites and reduce the required interaction time to perform the desired gate \cite{choi2014optimal, leung2018robust, milne2020phase, bentley2020numeric}. Users wishing to program at this level can use the Calibration Database to retrieve the relevant values such as the calibrated Rabi rate, the phase and frequency of the individual laser beams,  and several more. Our Symbolic Algebra then allows users to generate symbolic expressions using these parameters, to create \emph{e.g.} customized M\o lmer-S\o rensen interactions to implement entangling gates \cite{molmer1999multiparticle}. 

Users who prefer to program at the level of machine-specific gates instead of time-specific hardware instructions have the option to work in the native Gate Layer. This layer is enabled through a set of pre-defined pulse sequences in the Standard Library that implement standard gates such as individual $x$, $y$ and $z$-rotations and the $XX$ entangling gate. Combining these gates, users can construct their own version of a CNOT gate and customize gate parameters like rotation angles.

Lastly, the highest level of the programming stack is designed for users focusing purely on high-level quantum algorithms. Here, users can work directly with computational gates as provided by our Standard Library. The implementation of our CNOT gate, for example, can be traced back exactly in the manner we described. This example illustrates how these three key components, the Calibration Database, the Symbolic Algebra and the Standard Library come together to enable fully customizable quantum operations in our system.

\section{Technical details}

We will now take a look at the formal definition and actual syntax of the language, discuss the concept of language ``bindings'', and explain the language compilation process.

\subsection{Formal language definition}
Quala is defined in XML. We chose XML  due to its mature schema definition specification, its seamless integration with modern web security protocols, and the compatibility of its functional structure with our Symbolic Algebra needs. Additionally, defining the language elements in XML enables us to create an interface that is easily extensible to various high-level programming languages that have the ability to generate text-based output. These high-level language extensions are called language bindings.

Language bindings allow the users to write their code in a higher-level language of their choice. This can facilitate the process of writing programs with high-level programming paradigms such as loops and functions. The first binding we have implemented is in Python. We are planning on releasing bindings for Julia and Matlab in the future. All bindings use the same naming conventions and object relationships as in the XML language, and they all provide the ability to generate the required XML code at the end of the program definition. The Python binding, for example, provides a stand-alone Python module called \code{quala} that contains a 1:1 mapping of XML language elements to Python classes. All these classes are derived from a common base class that implements the translation of Python objects to XML tags. For example, the \code{NumericLiteral} element, which has the following definition in the language schema:



\begin{lstlisting}[language=XML]
<?xml version="1.0" encoding="UTF-8" ?>
<xs:schema xmlns:xs="http://www.w3.org/2001/XMLSchema"
  xmlns:qi="https://iqc.uwaterloo.ca/quantumion">

<xs:element name="NumericLiteral">
    <xs:annotation>
      <xs:documentation>A fixed numeric value with units. Can be any real number.</xs:documentation>
    </xs:annotation>
    <xs:complexType>
      <xs:simpleContent>
        <xs:extension base="xs:double">
          <xs:attribute name="units" type="xs:string"/>
        </xs:extension>
      </xs:simpleContent>
    </xs:complexType>
  </xs:element>
</xs:schema>
\end{lstlisting}

\noindent
can be used in Python through a class of the same name. Attributes and child tags are passed as arguments:


\begin{lstlisting}[language=mypython]
>>> import quala as ql
>>> value = ql.NumericLiteral(100, units="MHz")
>>> print(value.encode_xml())
.. <qi:NumericLiteral units="MHz">
     100
   </qi:NumericLiteral>
\end{lstlisting}

\noindent
The \code{encode\_xml} function is a special function that all objects in the binding languages share and which enables the translation between the binding object and the corresponding XML element.

Following the strict definitions set out in the language schema provides a natural way to enable the validation of user programs with linting tools such as \code{xmllint} and to flag syntax errors before the program reaches the compiler. This provides an additional layer of security as programs that do not  abide by the language definition cannot be passed through to the experiment.

\subsection{Elements of a Quala program}
\label{sec::program_components}

\begin{figure}[t!]
\includegraphics[scale=1.0]{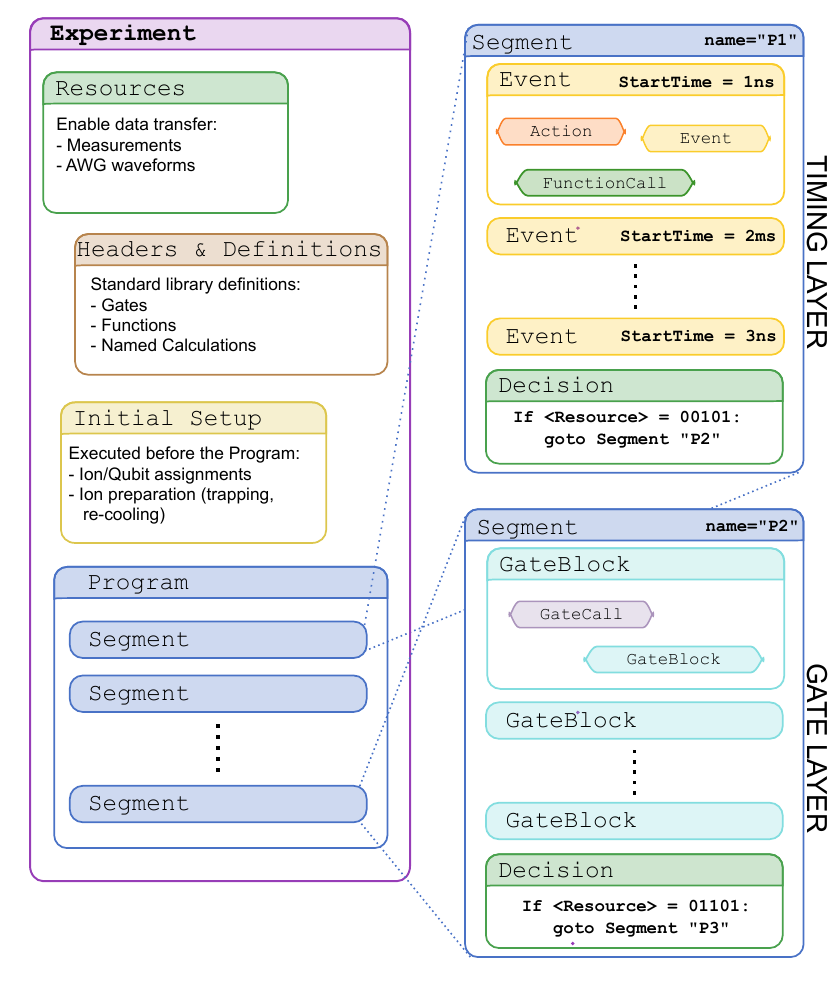}
\caption{Building blocks of a Quala program. This schematic illustrates the key elements of the language, represented here as rectangular boxes with background-colored titles, and their relationship to one another. For example, the \code{Experiment} element contains \code{Resources}, \code{Headers \& Definitions}, the \code{Initial Setup} and the \code{Program} element. The latter in turn is made up of a series of \code{Segment} elements, which can contain either \code{Event} or \code{GateBlock} elements as well as a \code{Decision} element at the end. For more information on the individual elements, see main text. Two full examples that utilize these elements are shown in \autoref{sec::examples}.}
\label{fig::program}
\end{figure}

The fundamental syntactical elements of our language and their relationships to each other are schematically shown in \autoref{fig::program}. The outermost component of every Quala program is called \code{Experiment}, which contains four different containers:

\begin{itemize}
    \item \code{Resources} specify containers for data storage such as measurement results or pulse waveform parameters.

    \item The \code{InitialSetup} contains a set of instructions for the experimental setup appear at the beginning of the experiment, such as desired number of ions/qubits and static oscillator frequencies.

    \item \code{Headers} and \code{Definitions} declare and define the Standard Library procedures are used in this experiment, if any. These include gates, functions and named calculations. We formally separate the declarations and definitions to allow for the possibility of encrypted function definitions.

    \item The \code{Program} contains the actual experimental instructions which are stored in individual \code{Segment} objects.
\end{itemize}

All experimental actions that are carried out during the user's program are defined within \code{Segment}s. There are two interchangeable ways of specifying the content of these elements, depending on whether the user would like to write programs in the Timing Layer with explicit timing specifications, or in the Gate Layer in which timing is implicit.

In the Timing Layer, a \code{Segment} contains a series of \code{Event} objects that specify which experimental actions are to take place at a given time. As such, every \code{Event} must define a \code{StartTime} that can be interpreted to be \emph{absolute}, with respect to the start of the experiment, or \emph{relative}, with respect to the \code{StartTime} of the previous \code{Event}. The other allowed elements within an \code{Event} are \code{Action}, \code{FunctionCall} and finally other \code{Event} tags. An \code{Action} specifies a direct and instantaneous experimental action, such as changing a DDS parameter or starting or stopping a photon counter. \code{FunctionCall} tags allow users to call pre-defined routines from the Standard Library, which are themselves made up of a series of \code{Event} tags.

In the Gate Layer on the other hand, a \code{Segment} contains a series of \code{GateBlock} objects that form a time-independent container for pre-defined gates that are scheduled to start at the same time. Users can access pre-defined gates in the Standard Library though \code{GateCall} objects. Similar to the \code{Event} tag, \code{GateBlock} tags can also be nested. The main difference between a \code{GateBlock} and an \code{Event} is that the time-dependence in the former is implicit. The introduction of \code{GateBlock}s is a mere convenience for users who are interested in circuit-level programming only; in principle every program can be fully expressed in a series of \code{Event}s, and these two containers may also be used interchangeably.

At the end of every \code{Segment}, users have the option of implementing a \code{Decision} block that can alter the execution flow of the program by jumping between different \code{Segment}s depending on the outcome of the conditions that are declared in the \code{Decision} object. Conditions are typically based on measurement outcomes, that are stored in \code{Resource} elements. While measurement outcomes in their raw form are photon counts, users can threshold the counts in real-time to obtain a 0 or 1 outcome for a qubit readout. It is also possible to combine multiple resources to compare the outcome of multiple qubits against multi-element bit strings.

Taking all these elements together, a complete (but largely empty) program in the Python binding may be constructed as follows:









\begin{lstlisting}[language=mypython]
import quala as ql

# Create a default InitialSetup element
setup = ql.InitialSetup(use_predefined="default")

# Generate a generic resource container
# (A measurement Action would fill it with data)
resources = ql.Resource(name="my_measurement")

# Generate an Segment with an empty Event,
# an empty GateBlock and a Decision
segment_1 = ql.Segment(
  ql.Event(
    # Note: Events must always specify a start time
    start_time=ql.NumericLiteral(0, "ns")
  ),
  ql.GateBlock(),
  ql.Decision(
    resource="my_measurement",
    conditions=[
      # If the outcome is "0", advance to "segment-2"
      ql.Condition("0", destination_segment="segment-2"),
      # If the outcome is "1", advance to "segment-3"
      ql.Condition("1", destination_segment="segment-3")
    ]
  )
)

# Generate two additional, empty Segments
segment_2 = ql.Segment(name="segment-2")
segment_3 = ql.Segment(name="segment-3")

# Combine all Segments within a Program
program = ql.Program(program_segments=[segment_1, segment_2, segment_3])

# Assemble the full Experiment
experiment = ql.Experiment(
  initial_setup=setup,
  resources=resources,
  program=program)
\end{lstlisting}

\noindent
Note that we have omitted the \code{Headers \& Definitions} element here, as those are inserted automatically when a program contains a call to the Standard Library in either the \code{Events} or the \code{GateBlocks}.

This abstract example highlights the relationships of the language elements to one another without making any assumptions about the actual experimental actions and the underlying physical hardware. To illustrate how our language can be used to implement realistic experiments, we provide two full-code examples in \autoref{sec::examples}.

\subsection{Language Compiler}
\label{sec::compiler}


Quala programs may contain abstract programming concepts such as looping constructs, functions, gates and symbolic calculations. Additionally, the decision logic and the ability to specify \code{Events} within \code{Events} allows for programs with several time-lines that are not directly translatable to hardware instructions. To bridge that gap, we have developed a sophisticated language compiler that reduces the functions and gates, flattens time-lines and solves the symbolic expressions. 

Compiling a Quala program follows three phases: frontend, middle, and backend processing, which is similar to the design philosophy of the GNU Compiler Collection (GCC) \cite{gcc-internals}.  At the first stage, the compiler parses the user's XML program into an object-oriented C++ data structure. This includes a validity check of the program through a linting tool and if errors are detected, the user may be given warnings or errors, depending on severity of problems.

The compiler-proper, or middle-end, performs a series of expansions that re-write the XML in successively less expressive forms.  This re-writing compiler process ensures that at all steps, the program is still a valid, equivalent XML program to what the user described.  The expressiveness is reduced at each step (meaning progressively simpler commands are used), leading to a simpler, albeit longer, program.

\begin{figure}[t!]
    \includegraphics[scale=1]{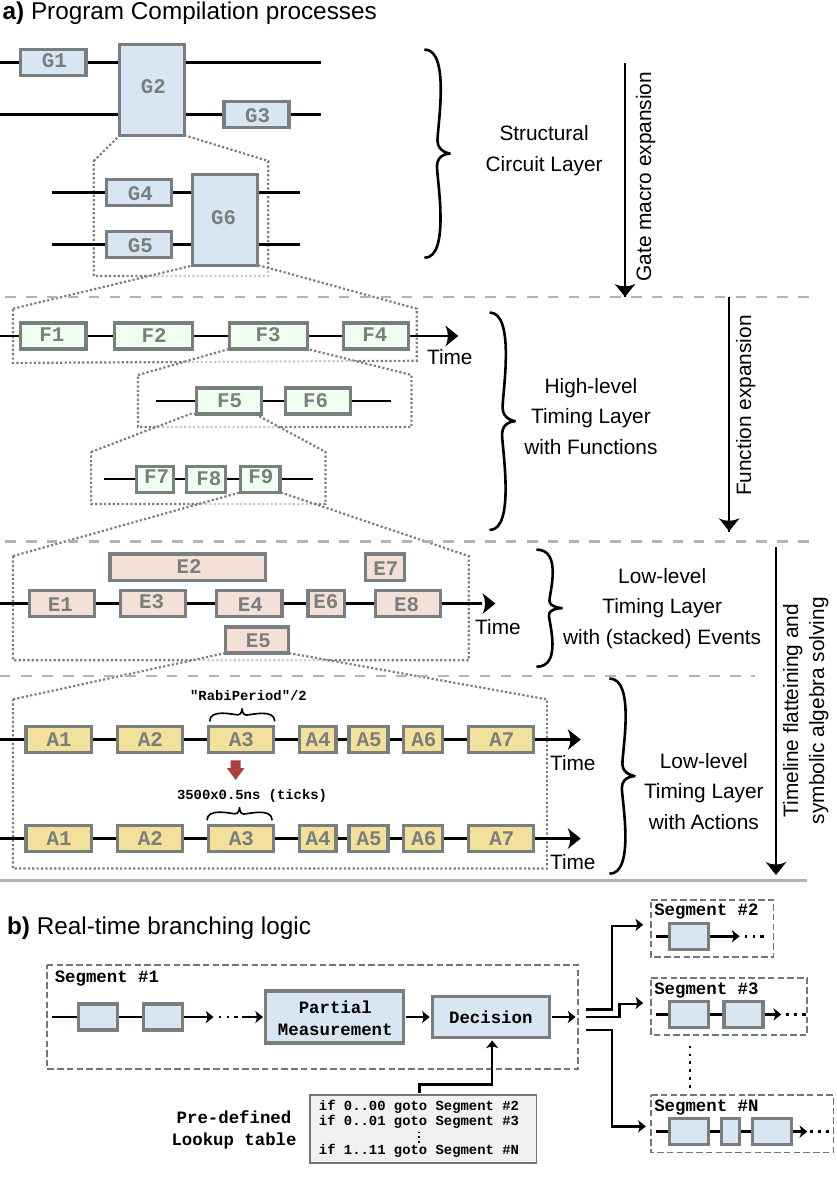}
    \caption{Language compiler and real-time decision logic. \textbf{a} illustrates the key steps of the compilation process that successively remove abstraction layers and simplify the program until it can be expressed as a flat time-line with direct experimental actions. \textbf{b} shows the schematic flow graph of the real-time branching logic. A pre-defined lookup table dictates the program execution flow.}
    \label{fig:branching-logic}
\end{figure}

The reduction steps are schematically shown in \autoref{fig:branching-logic} and can be summarized as follows:
\begin{itemize}
    \item The structural circuit model is decomposed by expanding \code{GateCalls} and \code{GateBlocks} into their definitions via macro expansion.
    \item The abstract Gate Layer is removed by expanding each \code{GateCalls} into function definitions.  The Standard Library provides this mapping.  This effectively \emph{solves} both the natural gate and the computational gate description as a pure Timing Layer  program.
    \item The \code{FunctionCalls} are expanded into their definitions from the Standard Library, or user functions, resulting only in Timing Layer \code{Events} and \code{Actions}.  Recursive calls are similarly solved.
    \item The resulting Timing Layer \code{Events} and \code{Actions} are \emph{flattened} by solving all relative start times and reordering \code{Events} into a single timeline.
    \item All symbolic expressions are solved with the latest database values (or specified historical value).  The result is simple numeric literal values.  Start times are now specified in units of $\unit[0.5]{ns}$ or `ticks'.
\item The composite actions, like \code{DDSAction}, are expanded into \emph{channelized} actions.  The result is simple, fully qualified names for each execution engine involved.  The resulting output program is viewable by the user in XML form as well.
\end{itemize}

At the last step of compilation, the so-called backend processing, converts the terminal XML form into a series of binary instructions, called `opcodes' that are used by our FPGA execution engine. 

\subsection{FPGA Execution Engine}

The language is designed to execute on FPGA hardware via the use of \emph{Execution Engines} and \emph{Action Cores}.  These modules reside inside the FPGA itself, and provide the programmability needed to implement language elements.  Action cores, as the name suggests, provide core logic for a particular action; examples include the phase accumulator and sine-lookup of a DDS core, or pulse counting of a photon measurement sensor.  Each type of Action element of a program controls an independent Action Core.

Execution engines are lightweight, synthetic microprocessors that provide all timing and interface to the user program.  Each user-controlled parameter, such as DDS phase, or amplitude, or photon-counter start/stop, is attached to a single, dedicated execution engine.  The small footprint of these engines allows many hundreds of units to be instantiated on a single FPGA chip.  With only slight changes in data word size, the execution engine module is the same regardless of the type of action core parameter being controlled.

The execution engine interfaces with an action core via standard Register Transfer Logic (RTL) methods.  The engines receive instructions in the form of operation codes (opcodes) from the terminal XML channelized output of the compiler.  A very simple opcode format allows small logic footprint, as shown in \autoref{tbl:opcodes}.  The primary opcode, \code{SETVALUE}, engages a change at a precise time.  The first generation execution engine does not contain internal state variables, but does allow for repetition loops.
\begin{table}
    \begin{tabular} { p{0.3\linewidth}  p{0.55\linewidth}  }
        \toprule
        \textbf{OPCODE} & \textbf{DESCRIPTION} \\
        \midrule
        \code{NOP} & No Operation \\
        \code{SETVALUE} $x$ & Set parameter value to $x$ \\
        \code{SETLOOP} $x$ & Sets the loop counter to $x$ \\
        \code{JNZ} $pc$ & Jump to instruction $pc$ if loop count is nonzero \\
        \code{JZ} $pc$ & Jump to instruction $pc$ if loop count is zero \\
        \code{DECLOOP} & Decrement the loop counter \\
        \code{GOTO} $pc$ & Unconditionally jump to $pc$ \\
        \code{BRANCHLUT} $m$, $t$ & Jump to instruction indicated by a look-up table $t$ based on measurement $m$  
\\
        \bottomrule
    \end{tabular}
    \caption{FPGA opcodes.  Each corresponds to a fundamental execution engine operation to implement user language features.}
    \label{tbl:opcodes}
\end{table}

Embedded in each opcode is a field indicating the time-delay that the instruction should be processed.  Unlike traditional microprocessors, which execute instructions at the next available cycle\footnote{Next-cycle execution is an oversimplification of pipelined, superscalar processors.}, the execution engine embeds a delay counter to every instruction.  These delay counters operate at the $\unit[2]{GHz}$ experiment rate, and define the precision timing of all system changes.  To overcome the timing closure demands of modern FPGA chips, the experiment clock rate of $\unit[2]{GHz}$ is interleaved in four clock phase-lanes, each operating at $\unit[500]{MHz}$.

\subsection{Decision Logic and Realtime Communications}

 
 The decision logic forms an integral part of our language and allows users to specify programs that execute certain blocks of code \emph{conditioned} on measurement outcomes. This is enabled through the \code{Decision} element (see \autoref{fig::program}) that can be placed at the end of each \code{Segment} to instruct the program which \code{Segment} to execute next, as illustrated in \autoref{fig:branching-logic}b.
 
 To support the decision logic in our hardware setup, each measurement result $m$ is broadcast over a low-latency Infiniband network.  Each FPGA module contains a copy of the decision block's lookup table $t$, generated by the language compiler.  When a \code{BRANCHLUT} $m$, $t$ instruction is executed, the execution engine performs a jump to the opcodes indicated by the appropriate destination segment based on that measurement $m$.

\section{Illustrative Examples}
\label{sec::examples}

Now we turn to two examples that highlight the core features and capabilities of our language both for digital and analog quantum programs. For the sake of brevity and clarity, we are omitting the machine-specific initial setup elements.

\subsection{5-Qubit error-correction code}


The five qubit code is a distance three error correction code that encodes a single logical qubit using 5 physical qubits \cite{laflamme1996perfect}. The fault-tolerant syndrome measurement circuit (FT-SMC) we use for our example uses two ancillary qubits, one to measure the syndrome bit and another to flag errors in the measurement process \cite{chao2018quantum}.

The code begins by performing a fault-tolerant syndrome measurement circuit (FT-SMC). If a fault is ever detected during this process, it aborts, performs a non-fault-tolerant syndrome measurement circuit (NFT-SMC), and uses the syndrome information to correct the fault. Overall, the algorithm can be broken down into the following high-level steps:  

\begin{enumerate}
    \item Prepare all qubits in computational ground state
    \item Run first FT-SMC and measure flag qubit 
    \begin{enumerate}
        \item If the flag is raised, interrupt, perform NFT-SMC, correct the fault, and terminate error correction
    \end{enumerate}
    \item Run second FT-SMC, measure flag qubit, repeat 2a)
    \item Run third FT-SMC, measure flag qubit, repeat 2a)
    \item Run fourth FT-SMC, measure flag qubit, repeat 2a)
    \item If no flag was raised perform error correction based on syndrome measurements
\end{enumerate}

The QuantumION language is uniquely suited to run programs of this form. We can express this algorithm through a series of \code{Segments}  linked together through \code{Decision} blocks. The gates for each syndrome measurement circuit and the individual measurements at the end can be packaged into a series of \code{GateBlocks}. To store the measurement results and make them available for the decision logic, we also need to create an empty \code{Resources} object at the beginning of our program.

With that, the $i$-th FT-SMC segment can be generated as follows:





\begin{lstlisting}[language=mypython]
import quala as ql

resources = ql.Resources(length=12)

def make_ft_smc(i: int) -> ql.Segment:
    """ Creates the i-th syndrome measurement circuit """
    ft_smc = ql.Segment(
      name=f"FT-SMC-{i}",
      ql.GateBlock(
        ql.GateCall("H", qubit=i, port="Target"),
        ql.GateCall("CX", qubit=(i, i+5), port=("Control", "Target")),
        # [more gates here]
        ql.GateCall("Measure", qubit=i+5, resource=resources[2*i])
        ql.GateCall("Measure", qubit=i+6, resource=resources[2*i+1])
      )
    )
    # Add the Decision block at the end
    decision_block = make_decisions(i)
    ft_smc.add(decision_block)

    return ft_smc
\end{lstlisting}

\noindent
The branching decision logic allows whole code blocks to be run (or not) conditioned on the result of a mid-circuit measurement.   

To implement the required \code{Decision} blocks, we need to generate a series of \code{Condition} objects that take in the measured state (0 or 1 in this example) and a ``destination'' \code{Segment}, to tell the compiler which \code{Segment} to execute next. These are referred to by their name. 



\begin{lstlisting}[language=mypython]
def make_decisions(i: int) -> ql.Decision: 
   """ Create the decision block for the i-th SMC segment  """
    segment_name = f"FT-SMC-{i}"
    next_segment_name = f"FT-SMC-{i+1}"

    decision = ql.Decision(
      resource=resources[2*i+1],
      conditions=[
        # If the outcome is 1, we want to jump to the non-fault-tolerant
        # SMC segment (not shown)
        ql.Condition(state=1, destination_segment="NFT-SMC"),
        # Otherwise, we want to continue to the next segment (definition shown above)
        ql.Condition(state=0, destination_segment=next_segment_name)
        ]
    )

    return decision
\end{lstlisting}

\noindent
Finally, we can create a series of these FT-SMC segments and string them together inside a \code{Program}.



\begin{lstlisting}[language=mypython]
program = ql.Program(
  # Create the FT-SMC segments
  segments=[make_ft_smc(i) for i in range(4)]
  # Add auxiliary correction segments and non-fault-tolerant segments (not shown)
  + aux_segments
)

experiment = ql.Experiment(program, resources)
\end{lstlisting}

\noindent
Thus, the QuantumION language can run the whole non-fault-tolerant syndrome measurement circuit with a single decision call from anywhere in the fault-tolerant circuit without any need for code duplication. The segment structure and branching decision logic remove the need for code duplication, and shuttling operations allow mid-circuit measurement without disturbing the rest of the computation. 

\subsection{Analog quantum simulation}

In this example, we generate an effective Ising interaction between two spins that can be described by the following Hamiltonian \cite{monroe2021programmable}

$$
H_\mathrm{Ising}(t) = \sum_{ij}^N J_{ij} \sigma_x^{(i)}\sigma_x^{(j)} + B_y(t) \sum_{i} \sigma_x^{(i)},
$$

Here, $N$ is the number of spin particles, $B_y(t)$ is the effective transverse magnetic field acting on each spin, and the $J_{ij}$ terms are the spin-spin coupling terms that depend on the drive parameters.
In order to generate this interaction in our setup, we need to encode three different waveforms on the laser beams that mediate the ion-ion interactions, two with static amplitudes and frequencies, and one with a time-dependent, decreasing amplitude (see \cite{richerme2013experimental} 
for more details).

To achieve this, we use a language element called \code{DDSAction} that allows for direct control over every DDSs (direct digital synthesizer, \emph{i.e.} RF signal generator) in the system. Like any other \code{Action} in our language, a \code{DDSAction} refers to an instantaneous, experimental instruction, and needs to be integrated into \code{Event} blocks to specify their start time and duration. We will define these actions first, and then demonstrate how they are embedded into \code{Events} and the overall program.

Every \code{DDSAction} takes as parameter a channel name, and can optionally take parameters such as the amplitude, frequency, absolute and relative phases, interpolation parameters and several more. For example, the static waveforms can be defined as follows:






\begin{lstlisting}[language=mypython]
import quala as ql

# Specify waveform frequencies relative to resonance
f0 = ql.NamedConstant("RamanCarrierResonanceFrequency")
f1 = f0 + ql.NumericLiteral(2, "MHz")
f2 = f0 - ql.NumericLiteral(2, "MHz")

# Set default amplitude
a0 = ql.NamedConstant("DefaultRamanIndividualDDSAmplitude")

ddsaction_1 = ql.DDSAction(
  channel="channels.aom.raman.individual1.dds0",
  amplitude=a0,
  frequency=f1
)

ddsaction_2 = ql.DDSAction(
  channel="channels.aom.raman.individual1.dds1",
  amplitude=a0,
  frequency=f2
)
\end{lstlisting}

\noindent
Where we have made use of the Calibration Database to import the system-specific Raman laser parameters through the \code{NamedConstant} element combined with our Symbolic Algebra. In order to program the time-dependent waveform, we can make use of the interpolation capabilities of our DDSs. For example, we can implement a linear sweep of the form $p(t) = p_1 + p_2*t$, between two amplitudes \code{a1} and \code{a2} over a fixed duration as follows:




  

\begin{lstlisting}[language=mypython]
# Define sweep duration
t_sweep = ql.NumericLiteral(10, "us")

# Set start and stop amplitudes (decreasing)
a1 = ql.NamedConstant("DefaultRamanIndividualDDSAmplitude")
a2 = a1 - ql.NumericLiteral(50, "mV")

ddsaction_3 = ql.DDSAction(
  channel="channels.aom.raman.individual1.dds2",
  frequency=f0,
  interp_type="polynomial",
  interp_p0=a1,
  interp_p1=(a2 - a1)/(t_sweep * ql.NamedConstant("DDSSampleClockFrequency")) 
  
)
\end{lstlisting}

\noindent
where we have used the Calibration Database parameter for the DDS clock frequency to calculate the interpolation parameter.

These three \code{DDSActions} form the heart of the analog quantum simulation routine. But before we can integrate them into a full quantum experiment, we first need to wrap \code{Events} around them that allow us to specify timing information. Two \code{Events} are needed: one for turning these three DDSs on, followed by another \code{Event} to turn them off. We can define those as follows:


\begin{lstlisting}[language=mypython]
# Define Ising interaction Events
ising_events = [
  # Event 1: Turn all DDSs on
  ql.Event(
    starttime=ql.StartTime(0, "ns"),
    event_items=[ddsaction_1, ddsaction_2, ddsaction_3]
  ),
  # Event 2: Turn all DDSs off
  ql.Event(
    starttime=ql.StartTime(t_sweep, stype="since-last-action"),
    event_items=[
      ql.DDSAction(
        channel="channels.aom.raman.individual1.dds0",
        amplitude=ql.NumericLiteral(0, "V")),
        ql.DDSAction(
        channel="channels.aom.raman.individual1.dds1",
        amplitude=ql.NumericLiteral(0, "V")),
        ql.DDSAction(
        channel="channels.aom.raman.individual1.dds2",
        amplitude=ql.NumericLiteral(0, "V")),
  )
]
\end{lstlisting}

\noindent

Note that we have set the \code{StartTime} of the second \code{Event} to begin at time \code{t\_sweep}, \emph{i.e.} our desired interaction time. To turn the interactions off, we simply set all DDS amplitudes to zero.

We can now put everything together into a full experiment. In addition to the Ising interaction, we need to include several preparation routines specific to our ion trap hardware. We have written all these routines up in the Standard Library, and we envision that users of other hardware platforms can add their hardware-specific routines too. In our setup, the full experimental protocol that we need to implement is as follows:

\begin{enumerate}
    \item Doppler Cooling
    \item Optical pumping to the ground state
    \item Sideband Cooling
    \item Global $\pi/2$ rotation
    \item Perform effective Ising interaction with the pre-defined \code{ising\_events}
    \item Global $\pi/2$ rotation
    \item Measurement
\end{enumerate}

\noindent
Since there is no decision logic in this example, we can package the entire experiment code into a single \code{Segment}. For most auxiliary routines, we can furthermore use the Standard Library. Since we want to include a measurement in this experimental sequence, we also need to declare a \code{Resource} that the measurement can be stored in. 




\begin{lstlisting}[language=mypython]
# Define measurement resource
r0 = ql.APDCounterResource()

segment = ql.Segment(
  segment_items=[
      ql.Event( # Step 1
          start_time=ql.NumericLiteral(0, "ns"),
          ql.FunctionCall("DopplerCooling", duration=ql.NumericLiteral(1, "ms"))
      ),
      ql.Event( # Step 2
          start_time=ql.NumericLiteral(0, "ns"),
          ql.FunctionCall("OpticalPumping", duration=ql.NumericLiteral(0.1, "ms"))
      ),
      ql.Event( # Step 3
          start_time=ql.NumericLiteral(0, "ns"),
          ql.FunctionCall("SidebandCooling")
      ),
      ql.GateBlock( # Step 4
        ql.GateCall("XPi/2", ion=0)
      ),
      # Step 5: Insert Ising interaction events (see below)
      *ising_events,
      ql.GateBlock( # Step 6
        ql.GateCall("XPi/2", ion=0)
      ),
      ql.Event( # Step 7
          start_time=ql.NumericLiteral(0, "ns"),
          ql.FunctionCall("GlobalReadout", duration=ql.NumericLiteral(0.5, "ms"), resource=r0)
      ),
  ]
)
\end{lstlisting}

\noindent
It is possible to combine both the Timing Layer's \code{Event} element with the Gate Layer's \code{GateBlock} in the same program. By default, the \code{start\_time} is taken relative to the last \code{Action} in the experiment, which in this case would be the last \code{Action} required to carry out the $\pi/2$ gates.

And finally, we can take all these elements together to construct the full \code{Experiment}:



\begin{lstlisting}[language=mypython]
experiment = ql.Experiment(
  program=ql.Program(segments=segment),
  resources=r0
)
\end{lstlisting}

The definitions of the auxiliary routines from the Standard Library are automatically packaged into the \code{Headers} and \code{Definition} elements, and the user does not need to specify those. Overall, this example implements the full experiment specification of an a analog quantum simulation experiment including all hardware controls required for its implementation.

\section{Conclusion and outlook}

We have presented a full-stack quantum programming language that is suitable for both hardware-agnostic circuit-level programming, as well as low-level hardware-specific timing layer programming. These two complementary views of a quantum program are integrated seamlessly in our language which offers an unprecedented degree of transparency into the analog nature of quantum hardware at a time where purely digital quantum computing is still largely dependent on sophisticated analog control design due to the high error budget of current quantum hardware. While this language was designed within the scope of the remote-access QuantumION platform, we believe the design philosophy and principles are applicable to a much broader range of hardware implementations and we hope that our work will provide a meaningful contribution to the quantum computing community and inspire discussions and collaborations on developing unified software frameworks for quantum programming.

\section{Acknowledgements}

We acknowledge support from the TQT (Transformative Quantum Technologies) research initiative at the University of Waterloo and the Natural Sciences and Engineering Research Council of
Canada (NSERC).

\bibliographystyle{ieeetr}

\end{document}